\documentclass[manuscript, review=false, anonymous=false, screen=true, authordraft=false, authorversion=true, nonacm=true]{acmart}
\usepackage[utf8]{inputenc}

\pdfoutput=1

\usepackage{tabularx}

\title{Theme Analysis of Political Facebook Ads in the 2021 Dutch General Election}

\author{Joren Vrancken}
\affiliation{
   \institution{Radboud University}
   \city{Nijmegen}
   \country{NL}
}
\email{jorenvrancken@gmail.com}
\acmDOI{}

\ccsdesc[500]{Human-centered computing~Social networking sites}
\ccsdesc[300]{Security and privacy~Social network security and privacy}
\ccsdesc[500]{Information systems~Online advertising}

\keywords{Microtargeting, Political Advertising, Political Campaigns, Facebook Ad Library, Elections, Dutch 2021 General Election}

\begin{abstract}
Social media platforms have been trying to be more transparent about the political ads they run on their platforms, because the Cambridge Analytica scandal revealed that political campaigns are using social media on a large scale. One such transparency effort is the Facebook Ad Library, a public repository of all political ads run on Facebook and Instagram. This library provides journalist and researchers with data to get a better understanding of political advertising and microtargeting on Facebook's platforms. Unfortunately, the Facebook Ad Library only provides estimates and basic information. This paper analyses political ads in more depth, by examining the themes that ads are about. We provide a method to match themes to political Facebook ads and we apply this method to analyse Facebook ad campaigns ran by Dutch political parties during the 2021 Dutch general election.
\end{abstract} 

\begin{document}

\newcommand{\inlineanonsupress}[1]{\emph{REDACTED FOR ANONYMITY}}
\begin{anonsuppress}
    \renewcommand{\inlineanonsupress}[1]{#1}
\end{anonsuppress}

\maketitle

\section{Introduction}
Social media has revolutionized the advertising business. Social media strive to provide a unique user experience to each user. The content one sees is tailored exactly to them. To accomplish this, social media platforms gather data from their users. They analyse this data to show content that keeps their users on the platform as long as possible. Social media platforms make money by showing ads to their users \cite{facebook_revenue}. Just as they try to tailor the content a user sees to their preferences, they also tailor the ads a user sees to their preferences. This way of delivering content and ads is called \emph{microtargeting}.

In traditional advertising, ads are seen by a large number of people (e.g. TV commercials). Many people that see these ads are not interested in their message and many others would have been interested anyway without seeing the ads. Microtargeting solves this problem. Instead of advertising to everyone that watches a particular TV channel or everybody that waits at a specific bus stop, social media platforms target ads to the people with high probability of being interested in them. Facebook provides advertisers with a large range of selectors to specify the right people \cite{facebook_business_targeting}. Facebook also helps advertisers find their ideal target audience. Facebook tracks the performance of ads to find the audiences that respond best to an advertiser's ads. For example, if an advertiser sells baby products, Facebook will let them target ads to women between the ages of 30 and 35, live in Amsterdam and are interested in products for young mothers. 

However, microtargeting is controversial, as it has become a tool for political campaigns to reach voters unnoticed. Borgesius et al. define microtargeting for political purpose as ``personalised communication that involves collecting information about people, and using that information to show them targeted political advertisements" \cite{threats_democracy}. Political microtargeting has become part of mainstream public debate after multiple media outlets reported that Cambridge Analytica, a political consultancy company, used data of tens of millions of Facebook users to microtarget political ads, without the users' consent \cite{nytimes_analytica_march} \cite{nytimes_analytica_april}. Microtargeting went unnoticed because journalists and researchers cannot get a transparent view of who political parties are targeting and what messages political parties are using, as the ads are only seen by the target audience.

As a response to this controversy, or to limit the chance that lawmakers adopt new strict laws, social media companies have amended their microtargeting practices. Twitter has decided to ban political advertising altogether \cite{twitter_ban_ads}. Facebook has created a public repository of all political ads that are run on their platforms (i.e. Facebook and Instagram), called the \emph{Facebook Ad Library} \cite{facebook_ad_library}. The library does not only provide the content of ads, but also metadata about the ads (e.g. how much was spent on the ad and when it was active). In theory, the goal of this library is to give better insights into the advertisements political campaigns are running. However, the library does have some transparency problems. For example, it does not show how much an ad cost, but instead only shows a range (e.g. between €2000 and €3000).

Political science scholars and political journalists are interested in the messaging of political parties, because the messaging of a political party shows which political themes a party is focusing on, what their stance is on those themes and what they want voters to know about those themes. Ads are an important part of this messaging. The Facebook Ad Library is a useful tool that researchers and journalists can use to analyse the messaging in political ads, however the Facebook Ad Library does not provide any metadata (e.g. what an ad is about) on the content. This makes it hard to analyse the content of the ads on a large scale. In this paper we present a technique to help researchers analyse the content of political ads (in the Facebook Ad Library), by matching themes to ads. We showcase this technique by analyzing the 2021 Dutch general elections.

First, we look at what data is available in the Facebook Ad library (\autoref{section: facebook ad library}). We then look at what Facebook ads look like and which elements are important for our analysis (\autoref{section: facebook ad details}). In \autoref{section: related work} we discuss related research. In \autoref{section: methodology} we describe the technique on how to match themes to ads and walk through each step. In \autoref{section: analysis} we apply this technique to the 2021 Dutch general elections and compare the results with existing voter research. We conclude in \autoref{section: conclusion} by discussing the methodology and the results. Finally, in \autoref{section: further research}, we look at some interesting leads to continue this research.

\subsection{The Facebook Ad Library}\label{section: facebook ad library}
The Facebook Ad Library is a public repository of ads published on Facebook and Instagram, released in 2019 \cite{facebook_ad_library_release}. It consists of two parts: a public website and an API \cite{facebook_ad_library_api}. The website can be used to manually search for ads. The API can be used by applications to automate searching and retrieving ads. 

\subsubsection{Information about ads}
The Facebook Ad Library provides a lot of information about each ad. The most important data the library provides for an ad is:
\begin{itemize}
    \item Start date: The date Facebook started showing an ad.
    
    \item End date\footnote{Not all ads have an end date, as active ads are also available in the Facebook Ad Library.}: The date Facebook stopped showing the ad.
    
    \item Spending: A range estimating the amount that is spent on the ad. The currency that was used in the payment is also provided.
    
    \item Impressions \cite{facebook_ad_impressions}: Impressions measure how many times an ad was shown to a user. Like spending, impressions are provided as a range estimating the amount.
    
    \item Estimated audience size (previously called potential reach) \cite{facebook_ad_potential_reach}: A range estimating how many users the ad could potentially be shown to.
    
    \item Demographic distribution: Facebook provides some basic demographic information of the distribution of impressions\footnote{The exact language Facebook uses is ``people reached by the ad", we assume this means the demographic information is about the impressions.}. They provide three demographic characteristics:
    \begin{itemize}
        \item Female/male ratio.
        
        \item Age groups\footnote{Facebook requires all users to be at least 13 years old.}: 13-17, 18-24, 25-34, 35-44, 45-54, 55-64, 65+.
        
        \item Regions: Large regions within a country. For example, in the Netherlands the regions are the twelve provinces.
    \end{itemize}

    \item Content: The actual text that appears in the ad. An ad has multiple text elements, see \autoref{section: facebook ad details} for more information. It should be noted that the images or videos of an ad are not directly accessible through the Facebook Ad Library API, but the API does provide a link to the ad including any image or video.
\end{itemize}

\subsubsection{Spending Tracker}
Besides information on ads, the library also provides information on the organisations and political parties behind the Facebook pages (Facebook ads are always linked to a Facebook page) that pay for the ads, in the so-called Spending Tracker \cite{facebook_ad_library_report}. This is a good resource to identify an exhaustive list of Facebook pages that are used by political parties to run ads.

\subsubsection{Limitations}
The Facebook Ad Library does have limitations \cite{library_pitfal}:
\begin{itemize}
    \item No exact numbers are given. The numeric data is given in ranges. For example, between €2000 and €3000 was spent on an ad or the ad has between 1000 and 2000 impressions.
    
    \item It is not clear when an ad was most active. For example, if an ad gained between 1000 and 2000 impressions and was active for 30 days, we do not know whether the ad was mostly shown on one day or evenly over the 30 days.
    
    \item There is a nuanced difference in the information that is available through the website and the API. For example, Facebook bundles similar ads on the library website to give a better estimation of the metrics. This information is not available through the API.
    
    
    \item It is not publicly known how Facebook decides who sees which ad. The Facebook Ad Library only provides basic information about whom an ad was shown to. This means that the Facebook Ad Library provides only half the picture. Because we do not know how this decision is made, we do not know whether it is an explicit decision by a political party to target certain people or whether it is an automated choice by the Facebook algorithm, or a combination.
\end{itemize}

\subsection{Anatomy of a Facebook Ad}\label{section: facebook ad details}
A Facebook Ad contains the following elements:
\begin{itemize}
    \item The top of the ad shows an header that informs the user that they are being shown an ad and which Facebook page has published and paid for the ad.
    
    \item \textbf{The creative body}: The main text of the ad.

    \item An image or video.
    
    \item Below the creative body and image we find a call to action. This section consists of the following elements: 
        \begin{itemize}
            \item \textbf{The creative link caption}: A link to a website of the advertiser.
            
            \item \textbf{The creative link title}: A title that is shown above the creative link caption.
            
            \item \textbf{The creative link description}: A description of the creative link caption.
        \end{itemize}
\end{itemize}

Not all of these elements need to be present in an ad (e.g. an ad can have an image but no creative body). It is also possible for ads to have multiple variants of one element (e.g. an ad with multiple creative bodies or multiple images). In this case the Facebook algorithm will pick the best combination of elements to show to a user. 

\section{Related Work}\label{section: related work}
The Facebook Ad Library has been used by researchers to get a better understanding of political advertising on social media. In 2020 Fowler et al. \cite{fowler_franz_martin_peskowitz_ridout_2021} used the Facebook Ad Library to compare political Facebook ads to more traditional political advertising (e.g. television commercials). Edelson et al. \cite{edelson2020security} use the Facebook Ad Library to find suspicious and malicious advertising practices. Schmøkel and Bossetta \cite{doi:10.1080/19331681.2021.1928579} provide tools to analyse the images in Facebook ads.

The Facebook Ad Library is not the only dataset that has been used to analyse ads on social media. ProPublica, in collaboration with other media outlets and researchers, published a dataset of Facebook ads (both political and non-political) that was gathered by volunteers installing a browser extension that saves all ads \cite{propublica-dataset}. Ortega \cite{lopez2021microtargeted} uses this dataset to analyse the use of negative campaigning with online political ads. Levi et al. \cite{10.1007/978-3-030-61841-4_7} use it to classify political ads.

Researchers have used other political texts (e.g. speeches) to analyse the themes that political parties and specific politicians focus on. Many researchers use topic modeling to extract themes from these political texts. Topic modeling is a class of natural language processing models to cluster words into groups of similar word, where each group should represent a theme. Latent Dirichlet Allocation (LDA), presented by Blei et al. \cite{lda}, is a widely used topic model. Topic modeling work best if the analysed documents are homogeneous. There are many models that implement topic modeling, each suitable for a specific type of text. For example, GSDMM (Yin et al. \cite{yin2014dirichlet}) is suitable for short texts. Political ads, however, come in many forms that range from just a few words to multi-paragraph articles. This makes topic modeling less suited for analyzing political ads. Instead of relying on topic models, we provide an alternative approach to match themes to political ads. 

 
\section{Methodology}\label{section: methodology}
The goal of our methodology is to match a theme (or multiple themes) to an ad. In other words, we want to categorize what an ad is about using a list of themes. To accomplish this we provide the following repetitive process:

\begin{enumerate}
    \item Create a list of themes (\autoref{section: methodology themes}).
    
    \item Obtain and pre-process the ad content (\autoref{section: methodology pre-processing});
    
    \item Create a list of relevant words for each theme (\autoref{section: methodology theme word lists});
    
    \item Match themes to ads using the word lists (\autoref{section: methodology matching});
    
    \item Update the word lists by using the matched ads (\autoref{section: methodolgy update lists});
    
    \item Go to step 3 (\autoref{section: methodogoly reiterate}).
\end{enumerate}

In this section we will go into detail about each step.

\subsection{Themes}\label{section: methodology themes}
We first need a list of themes to match to ads. These themes should cover the full public debate. We use existing research for this purpose. We base our themes on the codebooks from the Comparative Agendas Project (CAP) \cite{cap_mastercodebook}. These codebooks contain categories and sub-categories that cover every topic in the public debate and a description for each category.

As some themes are less prevalent in the current public debate and elections, it can be hard to distinguish between related themes. For example, it might be hard to distinguish between international affairs and foreign trade in ads. To solve this, we combine similar themes into one theme.

\subsection{Obtaining and Pre-processing Ad texts}\label{section: methodology pre-processing}
We use the Facebook Ad Library API to retrieve the ad data we need. We want to analyse the textual content of the ads, we are interested in three elements: the creative bodies, the creative link descriptions and the creative link titles. We will refer to these combined elements as the \emph{ad text}.

Before we can analyse the ad texts, we need to make sure they are all in the same format. We use common natural language processing for this: 
\begin{itemize}
    \item If an ad has multiple variants of an element (e.g. multiple possible creative bodies), combine them into a single text.
    \item Replace all non-ASCII characters (e.g. emojis and bold characters) with the nearest equivalent ASCII characters.
    \item Remove all words except for the nouns, proper nouns and adjectives.
    \item Normalize all words to their root form (e.g. ``cars" becomes ``car" and ``better" becomes ``good"). This is referred to as lemmatization.
\end{itemize}

To speed up the analysis, we also remove all duplicate texts, as political parties tend to reuse text.

\subsection{Creating Theme Word Lists}\label{section: methodology theme word lists}
For each theme we use a list of words that are relevant to that theme. We refer to these word lists as \emph{theme word lists}. We use the descriptions in the CAP codebooks for this purpose. For each theme, we (manually) add the relevant words from the corresponding CAP codebook category description to a list.

To make sure the theme word lists are all in the same format we pre-process them like the ad texts in \autoref{section: methodology pre-processing}.


\subsection{Matching Themes to Ads}\label{section: methodology matching}
Once we have the themes and their corresponding theme word lists we use the following algorithm to match themes to an ad:
\begin{enumerate}
    \item Compute the sizes of the intersections between each theme word list and the words in the ad text.
    
    \item If the largest intersection contains at most one common word, no theme is matched to the ad. This mean that ads do not have to correspond to a theme. This is necessary, because political parties regularly advertise about non-policy matters (e.g. events they are organizing and membership benefits).
    
    \item If the largest intersection contains more than one common word, match that theme to the ad.
    
    \item If an intersection contains more than five common words, match that theme to the ad (even if there is a larger intersection). This means that multiple themes can be matched to an ad. This is necessary as some advertisements are longer texts that cover multiple themes (e.g. a summary of the most important points of a party's policy plans). 
\end{enumerate}

\subsection{Updating the Theme Word Lists}\label{section: methodolgy update lists}
At this point we have ads that correspond to themes. We use these ads to improve our theme word lists by updating the theme word lists for a theme with words common in ads about that theme. For each theme, we compute the frequency of each word in the ads about that theme. We (manually) check the most common words for words that are relevant to the theme and (manually) add these to the theme word list.

Whether a word is relevant to a theme is subjective. This is ultimately up to the researcher creating the theme word list. We provide the following guidelines to aid in this decision:
\begin{itemize}

    \item When computing the most commonly used words in a set of ads, we will find many words that are commonly used in all ads (e.g. ``vote" and ``party" are common words in election ads) and words that are common in general (e.g. ``world" and ``year"). These need to be filtered out, as they do not correspond to a single theme.

    \item Ideally, a word should be in only one theme word list. If a word is relevant to multiple themes, it is probably better to not add it to any list as its meaning is too broad.
    
    \item Reading political ads helps getting a feeling for the way political ads are written and how words are used by political parties. This will help the researcher decide whether a word that has different meanings in different contexts should be added a theme word list. 
\end{itemize}

\subsection{Reiterating the Process}\label{section: methodogoly reiterate}
After an iteration of this process, we have improved theme word lists. We can use these improved lists to create an improved matching between themes and ads to find more relevant words for the theme word lists. Eventually, we come to a point where we do not find any more new, relevant words. At this point the process stops.

Some themes are more important in an election than others. As political parties advertise more about the themes that voters find important, there will be less unique, relevant words for some themes. This means that some theme word lists will be finished after fewer iterations than others. This also means that the theme word lists will not be of the same length.

\section{Analysis of 2021 Dutch General Election}\label{section: analysis}
In this section we will use the method described in \autoref{section: methodology} to analyse the themes used in political Facebook ads during the 2021 Dutch general election. This election was held from 15 to 17 March. We look at ads that ran between 1 September 2020 and 1 September 2021\footnote{We choose a full year to get a good picture of all ads leading up to the election and after the election. However, the vast majority of ads ran in the month leading up to the election.} and we focus on three parties that are seen as the winners of the election \cite{verkiezingen_nos} \cite{verkiezingen_volkskrant} \cite{verkiezingen_telegraaf} \cite{verkiezingen_nrc}: 
\begin{itemize}
    \item Democraten 66 (D66): The party that ended second in number of seats.
    \item Forum voor Democratie (FvD): The party that gained the most seats relative to their previous election results\footnote{FvD has since lost three of their eight seats, because three members split off.}.
    \item Volkspartij voor Vrijheid en Democratie (VVD): The party that won most seats.
\end{itemize}

These parties cover a broad political spectrum, D66 being a progressive, social liberal party, VVD a centre-right, conservative liberal party and FvD being a conservative, right-wing populist party \cite{dutch_parties}.

In \autoref{section: analysis themes} we look at the themes we used during this analysis. In \autoref{section: theme distribution results} we show the distribution of themes per party. We analyse these distributions through three lenses: we compare our results with existing voter research (\autoref{section: analysis most important themes}), we look at which parties got the most impressions for each theme (\autoref{section: analysis issue ownership}) and we look at the demographic make-up of the audience (\autoref{section: analysis demographics}). 

We published more detailed results (e.g. the results of all parties) on our website\footnote{\inlineanonsupress{\url{https://joren485.github.io/DutchPoliticalFacebookAdComparision/}}}. On this website we also published an analysis of the metadata of political Facebook Ads.   

\subsection{Themes}\label{section: analysis themes}
As described in \autoref{section: methodology themes} we create a list of themes using the CAP codebook of the Netherlands \cite{cap_netherlands}. This resulted in the following list of themes. The categories from the CAP codebook on which the themes are based are listed for each theme.
\begin{itemize}
    \item \textbf{Agriculture}: This theme covers all topics and debate relevant to agriculture, farming and livestock. This theme is based on category 4 (Agriculture and Fisheries).

    \item \textbf{Civil Rights}: This theme covers all topics and debate relevant to rights of citizens (e.g. discrimination and privacy). This theme is based on category 2 (Civil Rights). Category 2 also includes migration, which we split off into a new theme.
    
    \item \textbf{Climate}: This theme covers all topics and debate relevant to climate and the environment. This theme is based on categories 7 (Environment), 8 (Energy) and 21 (Public Lands).
    
    \item \textbf{Defense}: This theme covers all topics and debate relevant to defense and the military. This theme is based on category 16 (Defense). 
    
    \item \textbf{Economy}: This theme covers all topics and debate relevant to the economy (e.g. macroeconomic policies and commerce). This theme is based on categories 1 (Macroeconomics and taxes) and 15 (Domestic Commerce).
    
    \item \textbf{Education \& Culture}: This theme covers all topics and debate relevant to education, culture and religion. This theme is based on categories 6 (Education) and 17 (Technology).
    
    \item \textbf{Foreign Affairs}: This theme covers all topics and debate relevant to international relations and foreign affairs. This theme is based on categories 18 (Foreign Trade) and 19 (International Affairs).
    
    \item \textbf{Government}: This theme covers all topics and debate relevant to government services, government operations and public service. This theme is based on category 20 (Government Operations).
    
    \item \textbf{Healthcare}: This theme covers all topics and debate relevant to healthcare. This theme is based on category 3 (Healthcare).
    
    \item \textbf{Housing}: This theme covers all topics and debate relevant to housing and city planning. This theme is based on category 14 (Housing).

    \item \textbf{Law \& Order}: This theme covers all topics and debate relevant to general law, crime and jurisdiction. This theme is based on category 12 (Law and Crime).
    
    \item \textbf{Migration}: This theme covers all topics and debate relevant to immigration, emigration and refugees. This theme is split from category 2 (Civil Rights).
    
    \item \textbf{Social Welfare}: This theme covers all topics and debate relevant to social welfare (e.g. low-Income assistance). This theme is based on categories 5 (Labor) and 13 (Social Welfare).
    
    \item \textbf{Transportation}: This theme covers all topics and debate relevant to traffic, transportation and infrastructure. This theme is based on category 10 (Transportation).
\end{itemize}

The theme word lists we created be found at \inlineanonsupress{\url{https://github.com/joren485/DutchPoliticalFacebookAdComparision/tree/main/data/wordlists}}.

\subsection{Distribution of Themes in Ads}\label{section: theme distribution results}
Using the themes in \autoref{section: analysis themes}, we created theme word lists using our methodology and matched themes to the ads of D66, FvD and VVD. \autoref{table: theme distribution by ads} shows the distribution of themes in number of ads and \autoref{table: theme distribution by impressions} shows the distribution of themes of impressions\footnote{As Facebook does not provide a precise number of impressions, these numbers are based on averages.}.

\begin{table}[H]
\centering
\caption{The theme distribution by number of ads.}\label{table: theme distribution by ads}
\begin{tabular}[t]{l|l}
\textbf{Theme (D66)}     & \textbf{\%} \\ \hline
Education \& Culture &	33.88\%	\\
Climate &	26.82\%	\\
Healthcare &	14.42\%	\\
Housing &	9.68\%	\\
Civil Rights &	6.68\%	\\
Economy &	2.42\%	\\
Social Welfare &	2.23\%	\\
Transportation &	1.84\%	\\
Government &	0.77\%	\\
Law \& Order &	0.58\%	\\
Foreign Affairs &	0.39\%	\\
Agriculture &	0.29\%	\\
Defense &	0.0\%	\\
Migration &	0.0\%
\end{tabular}\hfill%
\begin{tabular}[t]{l|l}
\textbf{Theme (FvD)} & \textbf{\%} \\ \hline
Healthcare &	20.86\%	\\
Migration &	10.43\%	\\
Climate &	8.9\%	\\
Government &	8.9\%	\\
Economy &	7.67\%	\\
Civil Rights &	6.44\%	\\
Education \& Culture &	6.44\%	\\
Defense &	5.83\%	\\
Foreign Affairs &	5.21\%	\\
Housing &	5.21\%	\\
Social Welfare &	4.91\%	\\
Agriculture &	4.29\%	\\
Law \& Order &	3.07\%	\\
Transportation &	1.84\%
\end{tabular}\hfill%
\begin{tabular}[t]{l|ll}
\textbf{Theme (VVD)} & \textbf{\%} \\ \hline
Housing &	26.15\%	\\
Economy &	16.43\%	\\
Climate &	12.24\%	\\
Civil Rights &	8.77\%	\\
Healthcare &	8.45\%	\\
Social Welfare &	7.66\%	\\
Law \& Order &	6.95\%	\\
Education \& Culture &	3.87\%	\\
Transportation &	3.48\%	\\
Government &	2.76\%	\\
Defense &	1.82\%	\\
Migration &	0.71\%	\\
Agriculture &	0.39\%	\\
Foreign Affairs &	0.32\%
\end{tabular}\hfill%
\end{table}

\begin{table}[H]
\centering
\caption{The theme distribution by impressions.}\label{table: theme distribution by impressions}
\begin{tabular}[t]{l|l}
\textbf{Theme (D66)}     & \textbf{\%} \\ \hline
Climate &	30.84\%	\\
Education \& Culture &	27.31\%	\\
Healthcare &	15.66\%	\\
Housing &	11.44\%	\\
Civil Rights &	10.58\%	\\
Transportation &	1.32\%	\\
Economy &	1.09\%	\\
Social Welfare &	1.08\%	\\
Law \& Order &	0.27\%	\\
Government &	0.24\%	\\
Foreign Affairs &	0.09\%	\\
Agriculture &	0.08\%	\\
Defense &	0.0\%	\\
Migration &	0.0\%
\end{tabular}\hfill%
\begin{tabular}[t]{l|l}
\textbf{Theme (FvD)} & \textbf{\%} \\ \hline
Healthcare &	40.56\%	\\
Economy &	12.09\%	\\
Social Welfare &	7.48\%	\\
Civil Rights &	5.91\%	\\
Defense &	4.89\%	\\
Foreign Affairs &	4.81\%	\\
Housing &	4.38\%	\\
Climate &	3.93\%	\\
Migration &	3.31\%	\\
Agriculture &	2.97\%	\\
Education \& Culture &	2.96\%	\\
Government &	2.73\%	\\
Law \& Order &	2.5\%	\\
Transportation &	1.49\%
\end{tabular}\hfill%
\begin{tabular}[t]{l|ll}
\textbf{Theme (VVD)} & \textbf{\%}      \\ \hline
Civil Rights &	20.58\%	\\
Climate &	17.25\%	\\
Housing &	15.99\%	\\
Healthcare &	10.18\%	\\
Economy &	8.91\%	\\
Foreign Affairs &	8.72\%	\\
Law \& Order &	7.7\%	\\
Social Welfare &	5.01\%	\\
Education \& Culture &	2.08\%	\\
Defense &	0.97\%	\\
Transportation &	0.81\%	\\
Migration &	0.77\%	\\
Government &	0.73\%	\\
Agriculture &	0.29\%
\end{tabular}\hfill%
\end{table}

\subsubsection{Distribution of Matched and Non-matched Ads}
The calculations of \autoref{table: theme distribution by ads} and \autoref{table: theme distribution by impressions} only include ads that were matched to a theme. \autoref{table:matched percentage} shows what percentage of ads did not match any theme.

\begin{table}[H]
\centering
\caption{Which percentage of (the number of) ads is matched to at least one theme.}
\begin{tabular}{l|lll}
\textbf{Party} & \textbf{Number of Ads} & \textbf{Matched} & \textbf{Not matched} \\ \hline
D66 & 2640            & 34.68\%                      & 65.32\%                          \\
FvD & 672           & 41.48\%                      & 58.52\%                          \\
VVD & 6503           & 16.74\%                      & 83.26\%                         
\end{tabular}
\label{table:matched percentage}
\end{table}

There are multiple explanations why the matched percentage is relatively low for some parties:
\begin{itemize}
    \item Ads about organisational matters such as training sessions, conferences and becoming a member.
    
    \item Ads that focus on candidates or a broad message rather than on policy plans or themes. For example, in the election the VVD focused their campaign on the party leader and D66 focused on the message ``it is time for new leadership". Other parties focus more on their policy plans. For example, 79.59\% of ads by the Dutch labor party, Partij van de Arbeid (PvdA), were matched to a theme.
    
    \item Ads consisting of only a few words are harder to match to a specific theme.
\end{itemize}

\subsection{Most Important Themes According to Voters}\label{section: analysis most important themes}
Before the election, Nieuwsuur\footnote{A Dutch current affairs television show.} commissioned Ipsos\footnote{A market research company that regularly performs polling.} to do research into which themes are important to the constituency of each party \cite{ipsos_kiezersonderzoek}. The research includes the top five themes for each party:

\begin{tabularx}{0.9\textwidth}{ @{} X X X @{} }
  \textbf{D66}
  \begin{itemize}
    \item Climate \& Sustainability 
    \item Healthcare
    \item Education
    \item Housing
    \item Fighting COVID-19\footnote{Healthcare and Fighting COVID-19 from the Ipsos research are combined in the theme Healthcare in our research.}
  \end{itemize} &
  
  \textbf{FvD}
  \begin{itemize}
    \item Immigration 
    \item Healthcare
    \item Fighting COVID-19
    \item Norms and Values
    \item Economic Inequality
  \end{itemize} &
  
  \textbf{VVD}
  \begin{itemize}
    \item Healthcare
    \item Fighting COVID-19
    \item Employment
    \item Housing
    \item Safety
  \end{itemize}
\end{tabularx}

If we compare the top five themes according to the voters of each party with the top five themes in \autoref{section: theme distribution results}, we see almost a 1-to-1 overlap.

\begin{itemize}
    \item \textbf{D66}: The top five themes according to voters correspond the top four themes in ads. This shows us that D66 focuses its ads on themes that are important to their voters.
    
    \item \textbf{FvD}: Ads are highly focused on Healthcare, which their voters also care a lot about. However, the theme Climate is second in \% of ads, but not one of the most important themes according to their voters. We can also see this in \autoref{table: theme distribution by impressions}: Climate is in the 10th place of \% of impressions, showing us that these ads are (relatively) not performing well.
    
    \item \textbf{VVD}: The top five themes according to voters correspond to the top six themes in the ads. Like with FvD, the theme Climate is not one of the most important themes to the voters, but is prevalent in VVD ads.
\end{itemize}

\subsection{Most Impressions per Theme}\label{section: analysis issue ownership}
Issue ownership for a political party is defined as having the best solution for an issue according to voters. In this section we compare existing issue ownership research with which parties got the most impressions for each theme, to see whether the party with the best solutions is reaching the most voters with their messaging.

\autoref{table: top 3 parties per theme} shows the three parties that got the most impressions (and what percentage of the total impressions they got) for each theme. It should be noted, that this has an inherent bias to right leaning parties as they tend to advertise more than left leaning parties.

\begin{table}[ht]
\centering
\caption{The top three parties that got the most impressions for each theme.}\label{table: top 3 parties per theme}
\begin{tabular}{l|lll}
\textbf{Theme}     & \textbf{1st}   & \textbf{2nd}   & \textbf{3rd}   \\ \hline
Foreign Affairs	&	JA21 (29.91\%)	&	VVD (26.44\%)	&	FvD (26.44\%)		\\
Civil Rights	&	CDA (26.02\%)	&	PvdA (20.59\%)	&	SP (17.21\%)		\\
Defense	&	FvD (64.93\%)	&	CDA (11.03\%)	&	GL (10.14\%)		\\
Economy	&	CDA (29.61\%)	&	PvdA (22.96\%)	&	FvD (15.71\%)		\\
Healthcare	&	FvD (44.44\%)	&	GL (9.43\%)	&	PvdA (8.95\%)		\\
Housing	&	PvdA (23.71\%)	&	CDA (23.65\%)	&	SP (11.89\%)		\\
Law \& Order	&	JA21 (29.46\%)	&	CDA (22.31\%)	&	VVD (21.75\%)		\\
Climate	&	GL (23.36\%)	&	D66 (18.54\%)	&	PvdD (17.62\%)		\\
Agriculture	&	BBB (38.42\%)	&	CDA (26.83\%)	&	PvdD (20.17\%)		\\
Migration	&	JA21 (62.23\%)	&	FvD (17.43\%)	&	CDA (11.93\%)		\\
Education \& Culture	&	D66 (28.05\%)	&	CDA (20.85\%)	&	PvdA (13.83\%)		\\
Government	&	CDA (50.13\%)	&	SP (17.19\%)	&	FvD (13.82\%)		\\
Social Welfare	&	CDA (42.21\%)	&	PvdA (18.71\%)	&	SP (14.28\%)		\\
Transportation	&	GL (31.58\%)	&	FvD (16.11\%)	&	BBB (9.27\%)
\end{tabular}
\end{table}

On the 13th of February, I\&O Research\footnote{A Dutch research bureau researching social issues.}, commissioned by the Volkskrant\footnote{A Dutch newspaper.}, published voter research about the 2021 election\cite{io_kiezersonderzoek}. Section 2.3 of this report analyses issue ownership. In table 2.4 of the report, we can see the following issues are owned by D66, Fvd and VVD:

\begin{tabularx}{0.9\textwidth}{X X X}
  \textbf{D66}
  \begin{itemize}
    \item Education
    \item European Union
  \end{itemize} &
  
  \textbf{FvD}
  \begin{itemize}
    \item Immigration
    \item European Union
  \end{itemize} &
  
  \textbf{VVD}
  \begin{itemize}
    \item Safety
    \item Economy
    \item Employment
    \item Fighting COVID-19
    \item European Union
    \item Public Finances
    \item Counterterrorism
  \end{itemize}
\end{tabularx}

If we compare these issues with the data in \autoref{table: top 3 parties per theme} and \autoref{section: theme distribution results}, we make the following observations:

\begin{itemize}
    \item D66 owns Education and gets the most impressions for Education \& Culture.
    
    \item FvD advertises about issues that they do not own. For example, FvD advertises a lot about Healthcare and Climate, but does not own these issues.
    
    \item VVD owns multiple issues and focuses on most of them in their ads. However, if we look at \autoref{table: top 3 parties per theme}, we see that the VVD is only in the top three of impressions for two themes.
    
    \item The European Union issue from the I\&O report (which corresponds to the Foreign Affairs theme) does not get much attention in ads. For D66, FvD and VVD Foreign Affairs falls in the bottom three in \% of ads. However, the ads about Foreign Affairs do relatively well when we look at the \% of impressions. Especially the VVD gets many impressions and is second in number of impressions.
\end{itemize}

\subsection{Demographic Groups}\label{section: analysis demographics}
The demographic data that the Facebook Ad Library provides allows us to compute the distribution of ads among demographic groups. \autoref{appendix: demographic distribution per theme} shows the distribution of impressions (of all parties, not only D66, FvD and VVD) for each theme. The data is split into the three demographic types that Facebook provides: Female/male ratio, age groups and the twelve provinces. \autoref{appendix: demographic distribution per party} shows a similar table that splits the demographic data by party.

For comparison, the \emph{Population} column shows the population distribution of the demographic groups, based on data of the Statistics Netherlands\footnote{A Dutch governmental institution that gathers statistical information about the Netherlands. Centraal Bureau voor de Statistiek in Dutch.} \cite{cbs_age} \cite{cbs_kerncijfers}.

Using these tables we can make some interesting observations about the advertising strategy of the Dutch political parties:
\begin{itemize}
    \item There are some big swings in female/male distribution of the impressions. For example, nearly 60\% of all ads about migration are shown to men and nearly 60\% of all ads about social welfare are shown to women.

    \item As people between the ages of 13 and 17 are not allowed to vote, they are mostly excluded from the advertising.

    \item 65+ is the largest group by population, but this group is significantly under-represented in impressions for almost all themes. This may be because senior citizens are less active on social media.
    
    \item The distribution of impressions in provinces seems to be mostly inline with the population statistics. We only see a significant deviation from the population statistics in specific themes (e.g. Agriculture). 
\end{itemize}

\section{Discussion}\label{section: conclusion}
In \autoref{section: analysis most important themes} we see that themes that are important to voters are the same themes that political parties advertise about. This shows that our technique produces accurate results. It also shows that political parties prioritize themes that their constituencies find important.

As the theme word lists are updated with common words in ads, the word lists consist of commonly used words. This is advantageous, because political parties tend to use specific terms to convey their opinions. For example, when talking about climate change, a party that prioritizes the dangers of climate change will use different language than a party that is skeptic towards climate change. Only using pre-made lists of words or clustering similar words into groups (i.e. using topic modeling) might not include relevant words (and common misspellings), which will result in miss-matching themes to ads.

Different political parties use different strategies when it comes to writing ads. Some parties use many short ads each covering a single topic, others use large texts covering everything that is important to the party. This means that any technique to analyse the themes in ads should be agnostic to the length and format of the content of the ads. We accomplish this by only taking the intersection between the ad and the theme word lists into account and not looking at the length of the ad and the theme word list.

Our method does, however, have two inherent biases. Whether a word is relevant to a theme is subjective. This is an inherent bias towards the opinion of the researcher creating the theme word list. Whether a word in an ad is relevant to a theme is also dependent on the time the ad is active. The problems that are relevant to a certain theme are different each election, this means that there is a bias towards the current political issues in the theme word lists.

As discussed in \autoref{section: methodolgy update lists}, someone needs to decide whether a word appearing in an ad is relevant to a theme or not. Unfortunately, this creates a lot of manual work for the researcher. However, as we only look at words that are common in ads, the method is still scalable. For example, we analysed more than thirty thousand ads for this research.

Although the Facebook Ad Library gives a lot of insights into the ads that are run on the platforms of Facebook, we still only see part of the picture. Facebook only provides estimates of the performance of an ad. These estimates make it possible to analyse the advertising performance of a theme or political party. However, it does not help us answer a maybe more important question: How are political parties using opaque online tools (such as microtargeting) to reach voters?

\section{Future Work}\label{section: further research}
\begin{itemize}
    \item In this research we focused on the text contained in political ads. However, many ads also include an image or video. Some ads only consist of an image or video and do not have any text. These images and videos should also be analysed to improve the theme matching.
    
    \item In \autoref{section: analysis} we focused on the distribution of themes in ads by political parties. It would be interesting to look more deeply into the way political parties talk about theses themes, for example, sentiment analysis.
    
    \item As Edelson et al. \cite{edelson2020security} shows, we can see which seemingly separate advertisers publish closely related (or identical) ads. We might be able to find parties that are collaborating in their ad campaigns.
    
\end{itemize}

\begin{acks}
We would like to sincerely thank Frederik Zuiderveen Borgesius and Tom Dobber. They have provided invaluable guidance, feedback and advice during this research. We would also like to sincerely thank everybody that proofread this paper.
\end{acks}

\bibliography{references}

\appendix

\section{Demographic Distribution Tables}\label{appendix: demographic distribution per theme}

\begin{table}[H]
\caption{The demographic distribution of impressions per theme.}
\begin{tabular}{ll|lllllll}
\hline
\textbf{Dem.} &
  \textit{\textbf{Pop.}} &
  \rotatebox{60}{\textbf{Agriculture}} &
  \rotatebox{60}{\textbf{\begin{tabular}[c]{@{}l@{}}Civil\\ Rights\end{tabular}}} &
  \rotatebox{60}{\textbf{Climate}} &
  \rotatebox{60}{\textbf{Defense}} &
  \rotatebox{60}{\textbf{Economy}} &
  \rotatebox{60}{\textbf{\begin{tabular}[c]{@{}l@{}}Education\\ \& Culture\end{tabular}}} &
  \rotatebox{60}{\textbf{\begin{tabular}[c]{@{}l@{}}Foreign\\ Affairs\end{tabular}}} \\ \hline
\textbf{Female}        & \textit{50.29\%} & 47.08\% & 58.99\% & 50.31\% & 41.09\% & 54.09\% & 54.76\% & 45.77\% \\
\textbf{Male}          & \textit{49.71\%} & 52.92\% & 41.01\% & 49.69\% & 58.91\% & 45.91\% & 45.24\% & 54.23\% \\ \hline
\textbf{13-17}         & \textit{6.47\%}  & 0.09\%  & 0.22\%  & 0.06\%  & 0.01\%  & 0.33\%  & 0.14\%  & 0.04\%  \\
\textbf{18-24}         & \textit{10.17\%} & 17.39\% & 16.09\% & 19.37\% & 13.34\% & 13.31\% & 21.82\% & 11.78\% \\
\textbf{25-34}         & \textit{14.92\%} & 25.24\% & 19.45\% & 24.21\% & 21.60\% & 20.27\% & 21.83\% & 20.23\% \\
\textbf{35-44}         & \textit{13.78\%} & 16.67\% & 14.01\% & 15.03\% & 19.36\% & 15.27\% & 13.53\% & 17.01\% \\
\textbf{45-54}         & \textit{15.96\%} & 15.86\% & 14.77\% & 19.55\% & 18.94\% & 17.75\% & 14.62\% & 18.65\% \\
\textbf{55-64}         & \textit{15.86\%} & 13.51\% & 15.60\% & 13.81\% & 15.37\% & 16.07\% & 13.75\% & 17.58\% \\
\textbf{65+}           & \textit{22.84\%} & 11.24\% & 19.86\% & 7.98\%  & 11.39\% & 17.00\% & 14.31\% & 14.72\% \\ \hline
\textbf{Drenthe}       & \textit{2.83\%}  & 6.00\%  & 3.26\%  & 4.03\%  & 4.01\%  & 4.14\%  & 3.48\%  & 4.22\%  \\
\textbf{Flevoland}     & \textit{2.45\%}  & 2.01\%  & 2.30\%  & 2.26\%  & 2.14\%  & 2.41\%  & 2.48\%  & 2.46\%  \\
\textbf{Friesland}     & \textit{3.73\%}  & 6.04\%  & 3.76\%  & 4.74\%  & 4.69\%  & 4.55\%  & 3.94\%  & 4.82\%  \\
\textbf{Gelderland}    & \textit{12.00\%} & 14.20\% & 11.45\% & 12.33\% & 11.90\% & 11.90\% & 13.46\% & 12.14\% \\
\textbf{Groningen}     & \textit{3.36\%}  & 4.62\%  & 3.62\%  & 4.12\%  & 3.96\%  & 4.96\%  & 3.83\%  & 4.20\%  \\
\textbf{Limburg}       & \textit{6.39\%}  & 7.18\%  & 8.31\%  & 6.54\%  & 7.84\%  & 7.71\%  & 6.36\%  & 7.82\%  \\
\textbf{Noord-Brabant} & \textit{14.73\%} & 14.34\% & 15.30\% & 13.81\% & 13.83\% & 14.89\% & 13.57\% & 13.75\% \\
\textbf{Noord-Holland} & \textit{16.53\%} & 9.59\%  & 14.64\% & 15.55\% & 17.31\% & 12.78\% & 13.83\% & 13.24\% \\
\textbf{Overijssel}    & \textit{6.68\%}  & 12.46\% & 7.50\%  & 7.52\%  & 7.71\%  & 8.58\%  & 8.53\%  & 9.66\%  \\
\textbf{Utrecht}       & \textit{7.79\%}  & 6.34\%  & 5.98\%  & 6.59\%  & 5.24\%  & 6.18\%  & 6.43\%  & 5.09\%  \\
\textbf{Zeeland}       & \textit{2.21\%}  & 3.77\%  & 2.76\%  & 2.78\%  & 2.46\%  & 2.73\%  & 3.66\%  & 2.64\%  \\
\textbf{Zuid-Holland}  & \textit{21.32\%} & 13.46\% & 21.11\% & 19.72\% & 18.92\% & 19.17\% & 20.44\% & 19.95\% \\ \hline
\end{tabular}
\end{table}

\begin{table}[H]
\raggedright
\caption{The demographic distribution of impressions per theme.}
\begin{tabular}{ll|lllllll}
\hline
\textbf{Dem.} &
  \textit{\textbf{Pop.}} &
  \rotatebox{60}{\textbf{Government}} &
  \rotatebox{60}{\textbf{Healthcare}} &
  \rotatebox{60}{\textbf{Housing}} &
  \rotatebox{60}{\textbf{Law \& Order}} &
  \rotatebox{60}{\textbf{Migration}} &
  \rotatebox{60}{\textbf{Social Welfare}} &
  \rotatebox{60}{\textbf{Transportation}} \\ \hline
\textbf{Female}        & \textit{50.29\%} & 51.78\% & 52.82\% & 54.89\% & 47.17\% & 41.48\% & 58.80\% & 47.21\% \\
\textbf{Male}          & \textit{49.71\%} & 48.22\% & 47.18\% & 45.11\% & 52.83\% & 58.52\% & 41.20\% & 52.79\% \\ \hline
\textbf{13-17}         & \textit{6.47\%}  & 0.10\%  & 0.03\%  & 0.10\%  & 0.06\%  & 0.02\%  & 0.35\%  & 0.18\%  \\
\textbf{18-24}         & \textit{10.17\%} & 9.55\%  & 16.15\% & 17.02\% & 17.93\% & 9.88\%  & 11.56\% & 13.25\% \\
\textbf{25-34}         & \textit{14.92\%} & 16.51\% & 22.55\% & 24.76\% & 22.14\% & 16.42\% & 16.92\% & 20.05\% \\
\textbf{35-44}         & \textit{13.78\%} & 13.88\% & 15.81\% & 14.75\% & 14.37\% & 14.12\% & 13.26\% & 14.01\% \\
\textbf{45-54}         & \textit{15.96\%} & 16.24\% & 16.70\% & 13.83\% & 15.57\% & 17.11\% & 14.91\% & 21.32\% \\
\textbf{55-64}         & \textit{15.86\%} & 19.79\% & 15.38\% & 14.02\% & 15.30\% & 20.01\% & 18.39\% & 17.84\% \\
\textbf{65+}           & \textit{22.84\%} & 23.93\% & 13.37\% & 15.53\% & 14.62\% & 22.44\% & 24.62\% & 13.36\% \\ \hline
\textbf{Drenthe}       & \textit{2.83\%}  & 4.07\%  & 3.99\%  & 3.49\%  & 3.83\%  & 4.26\%  & 3.65\%  & 3.74\%  \\
\textbf{Flevoland}     & \textit{2.45\%}  & 2.35\%  & 2.45\%  & 2.04\%  & 2.31\%  & 2.47\%  & 2.41\%  & 2.13\%  \\
\textbf{Friesland}     & \textit{3.73\%}  & 3.66\%  & 4.67\%  & 4.18\%  & 4.21\%  & 5.05\%  & 4.28\%  & 4.40\%  \\
\textbf{Gelderland}    & \textit{12.00\%} & 12.30\% & 12.23\% & 10.37\% & 11.73\% & 12.03\% & 12.38\% & 12.70\% \\
\textbf{Groningen}     & \textit{3.36\%}  & 4.02\%  & 4.17\%  & 4.23\%  & 3.84\%  & 4.32\%  & 3.81\%  & 3.57\%  \\
\textbf{Limburg}       & \textit{6.39\%}  & 6.97\%  & 7.69\%  & 8.18\%  & 7.62\%  & 8.62\%  & 7.78\%  & 5.47\%  \\
\textbf{Noord-Brabant} & \textit{14.73\%} & 13.50\% & 14.00\% & 13.97\% & 14.42\% & 13.86\% & 15.74\% & 9.80\%  \\
\textbf{Noord-Holland} & \textit{16.53\%} & 11.82\% & 13.83\% & 13.48\% & 13.02\% & 12.95\% & 13.21\% & 11.45\% \\
\textbf{Overijssel}    & \textit{6.68\%}  & 7.28\%  & 8.12\%  & 9.53\%  & 9.59\%  & 8.63\%  & 7.98\%  & 13.40\% \\
\textbf{Utrecht}       & \textit{7.79\%}  & 6.76\%  & 5.78\%  & 6.70\%  & 6.05\%  & 5.30\%  & 6.36\%  & 13.93\% \\
\textbf{Zeeland}       & \textit{2.21\%}  & 5.36\%  & 2.92\%  & 2.77\%  & 2.88\%  & 2.49\%  & 3.66\%  & 3.53\%  \\
\textbf{Zuid-Holland}  & \textit{21.32\%} & 21.91\% & 20.16\% & 21.06\% & 20.50\% & 20.01\% & 18.75\% & 15.87\% \\ \hline
\end{tabular}
\end{table}

\section{Distribution of Impressions per Demographic by Party}\label{appendix: demographic distribution per party}

\begin{table}[H]
\caption{The demographic distribution of impressions for D66, FvD and VVD.}
\begin{tabular}{@{}ll|lll@{}}
\toprule
\textbf{Demographic} & \textit{\textbf{Population}} & \textbf{D66} & \textbf{FvD} & \textbf{VVD} \\ \midrule
\textbf{Female}        & \textit{50.29\%} & 53.33\% & 41.47\% & 48.28\% \\
\textbf{Male}          & \textit{49.71\%} & 46.67\% & 58.53\% & 51.72\% \\ \midrule
\textbf{13-17}         & \textit{6.47\%}  & 0.04\%  & 0.03\%  & 0.04\%  \\
\textbf{18-24}         & \textit{10.17\%} & 22.49\% & 19.31\% & 16.29\% \\
\textbf{25-34}         & \textit{14.92\%} & 24.53\% & 28.32\% & 23.48\% \\
\textbf{35-44}         & \textit{13.78\%} & 14.69\% & 15.03\% & 18.32\% \\
\textbf{45-54}         & \textit{15.96\%} & 14.38\% & 15.30\% & 18.31\% \\
\textbf{55-64}         & \textit{15.86\%} & 12.68\% & 12.71\% & 13.79\% \\
\textbf{65+}           & \textit{22.84\%} & 11.19\% & 9.31\%  & 9.79\%  \\ \midrule
\textbf{Drenthe}       & \textit{2.83\%}  & 3.54\%  & 4.21\%  & 3.82\%  \\
\textbf{Flevoland}     & \textit{2.45\%}  & 2.38\%  & 2.33\%  & 2.02\%  \\
\textbf{Friesland}     & \textit{3.73\%}  & 4.36\%  & 5.07\%  & 3.45\%  \\
\textbf{Gelderland}    & \textit{12.00\%} & 12.22\% & 11.81\% & 11.60\% \\
\textbf{Groningen}     & \textit{3.36\%}  & 4.11\%  & 4.20\%  & 3.10\%  \\
\textbf{Limburg}       & \textit{6.39\%}  & 7.05\%  & 8.12\%  & 6.61\%  \\
\textbf{Noord-Brabant} & \textit{14.73\%} & 14.95\% & 15.41\% & 16.84\% \\
\textbf{Noord-Holland} & \textit{16.53\%} & 14.38\% & 12.51\% & 15.36\% \\
\textbf{Overijssel}    & \textit{6.68\%}  & 7.52\%  & 8.50\%  & 7.16\%  \\
\textbf{Utrecht}       & \textit{7.79\%}  & 6.71\%  & 4.83\%  & 7.06\%  \\
\textbf{Zeeland}       & \textit{2.21\%}  & 2.40\%  & 2.74\%  & 2.09\%  \\
\textbf{Zuid-Holland}  & \textit{21.32\%} & 20.38\% & 20.26\% & 20.90\% \\ \bottomrule
\end{tabular}
\end{table}

\end{document}